\documentclass[epj]{svjour}
\usepackage{amssymb}
\usepackage{amsmath}
\usepackage{graphicx}
\usepackage{sidecap}
\usepackage{epstopdf}
\usepackage{bm}%
\usepackage{gensymb}
\usepackage{soul}
\usepackage{sidecap}
\usepackage{bbm}
\usepackage[normalem]{ulem}
\usepackage {times}

\newcommand{\bra}[1]{\ensuremath{\langle #1 |}}
\newcommand{\ket}[1]{\ensuremath{| #1 \rangle}}

\newcommand{\be}{\begin{equation}}
\newcommand{\ee}{\end{equation}}
\newcommand{\bk}{{{\bf{k}}}}
\newcommand{\bq}{{{\bf{q}}}}

\newcommand{\br}{{{\bf{r}}}}

\newcommand{\bg}{{{\bf{g}}}}

\newcommand{\bea}{\begin{eqnarray}}
\newcommand{\eea}{\end{eqnarray}}

\newcommand{\bS}{{\bf S}}

\newcommand{\bd}{\begin{displaymath}}
\newcommand{\ed}{\end{displaymath}}
\newcommand{\ba}{\begin{array}}
\newcommand{\ea}{\end{array}}
\newcommand{\bi}{\begin{itemize}}
\newcommand{\ei}{\end{itemize}}
\newcommand{\bc}{\begin{center}}
\newcommand{\ec}{\end{center}}
\newcommand{\bfl}{\begin{flushleft}}
\newcommand{\efl}{\end{flushleft}}
\newcommand{\bfr}{\begin{flushright}}
\newcommand{\efr}{\end{flushright}}
\newcommand{\no}{\nonumber}

\newcommand{\mi}{\rm i}

\newcommand\redsout{\bgroup\markoverwith{\textcolor{red}{\rule[0.5ex]{2pt}{0.4pt}}}\ULon}
\newcommand\bluesout{\bgroup\markoverwith{\textcolor{blue}{\rule[0.5ex]{2pt}{0.4pt}}}\ULon}

\newcommand{\bl}{\begin{aligned}}
\newcommand{\el}{\end{aligned}}

\def\ket#1{\left\vert #1 \right\rangle}
\def\br{{\bf r}}
\def\bk{{\bf k}} \def\bq{{\bf q}}  
\def\bg{{\bf g}}  \def\bd{{\bf d}}  \def\bS{{\bf S}}
 \def\bS{{\bf S}}

\def\6{\partial}

\def\bra{\langle}
\def\ket{\rangle}
\def\={\!\!\!&=&\!\!\!}
\def\+{\!\!\!&&\!\!\!+~}
\def\-{\!\!\!&&\!\!\!-~}

 \newcommand\redout{\bgroup\markoverwith{\textcolor{red}{\rule[.5ex]{2pt}{0.4pt}}}\ULon}

\usepackage[colorlinks=true,citecolor=blue]{hyperref}
\hypersetup{colorlinks=true,citecolor=blue,linkcolor=red,urlcolor=blue}

\title{Momentum space imaging of nonsymmorphic superconductors with locally broken inversion symmetry}
\author{Mehdi Biderang\inst{1,2,3} \and Mohammad-Hossein Zare\inst{4} \and Alireza Akbari\inst{2,5,6}}

\institute{ 
	 \inst{1} Department of Physics and Astronomy, University of Manitoba, Winnipeg R3T 2N2, Canada 
	 \mail {mehdi.biderang@umanitoba.ca, and akbari@postech.ac.kr}
\\       
	\inst{2} Asia Pacific Center for Theoretical Physics, Pohang, Gyeongbuk 790-784, Korea
\\
    \inst{3} Manitoba Quantum Institute, University of Manitoba, Winnipeg R3T 2N2, Canada	  
\\
	 \inst{4} Department of Physics, Faculty of Science, Qom University of Technology, Qom 37181-46645, Iran
\\     
\inst{5} Max Planck POSTECH Center for Complex Phase Materials,  and Dept. of Physics, POSTECH, Pohang 790-784, Korea
\\
\inst{6} Max Planck Institute for the Chemical Physics of Solids, D-01187 Dresden, Germany
}
\abstract{
The failure of spatial inversion symmetry in noncentrosymmetric materials introduces  two different types of spin-independent and spin-dependent electron hopping.
The spin-dependent term can be translated into  a quasi-spin-orbit coupling and may affect the electronic structure. 
In the locally noncentrosymmetric crystals, the presence of a sublattice degree of freedom generates a distinction between the inter- and intra-sublattice hopping integrals.
The spin-dependent part of the former (latter), which is even (odd) under parity, is called symmetric (antisymmetric) quasi-spin-orbit coupling. 
Here, we show the consequences of  such quasi-spin-orbit couplings on the electronic band structure, and
 study their characteristic features via the quasiparticle interference method. 
We extend our discussions to a  realistic class of materials, known as transition metal oxides.
}
\begin{document}
\maketitle
\section{Introduction}
The discovery of superconductivity in heavy fermion CePt$_3$Si has begun a new era in the topic of superconductivity in systems without spatial inversion symmetry~\cite{Sigrist_PRL_2004,Sigrist_Agterberg_PRL_2004,Yanase_Sigrist_JPJS_2008}.
The lack of inversion center affects the electronic attributions 
 through a symmetry-specified antisymmetric spin-orbit coupling~\cite{Fischer_PRB_2011}, which is the microscopic key  ingredient to understand the physics of noncentrosymmetric (NCS) systems.
This entangles the orbital and spin of the electrons and generates some features in the superconducting order parameter, as a mixture of even and odd parity Cooper pairings~\cite{Smidman_Rep_Prog_2017}.
Consequently, many interesting properties and promising industrial applications emerge, such as nontrivial topological properties, Majorana fermions, and quantum computation facilities~\cite{Sato_PRB_2009,Schnyder_PRB_2012,Schnyder_PRL_2013}.
%

Recently, a great deal of attention has been paid to a group of materials that, while having a spatial inversion symmetry in global, locally lack this type of symmetry~\cite{Maruyama_JPSJ_2012,Schnyder_PRB_2013,Yoshida_JPSJ_2014,Akbari_epl_2014,Yoshida_PRL_2015}.
This property may be due to stacking faults in the crystal structure, where the global inversion symmetry, as a result of the random distribution of structural faults, is preserved~\cite{Yanase_PRB_2018}.
Moreover, lack of some crystal properties such as rotational or translational symmetries may results in locally broken of inversion symmetry~\cite{Yanase_JPSJ_2010}.
In most cases, the sublattice degree of freedom associated with an antiferromagnetic order causes the doubling of ordinary unit cell~\cite{Fischer_PRB_2011}, and leads to leave the structure invariant under $D^{}_{4h}$ subgroups.
An objective example  is a layered structure with tetragonal symmetry, found in  some transition metal oxides (TMOs),
like provskite irridates in Ruddlesden-Popper phases (Sr$_{n+1}$Ir$_{n}$O$_{3n+1}$)~\cite{Cao_RPP_2018,BJ_Kim_Annurev_2019}. %
The sub-units of them are rotated octahedra, consisting of six oxygen ions encountering an Ir atom. 
The staggered rotations of each sub-unit along the $[001]$ direction is resulted from swivel of adjacent octahedra in the $xy$ plane in the opposite direction.
Consequently, the in-plane Ir-O-Ir bonds deviate from $180^{\circ}$, leading to breaking of spatial inversion symmetry on each individual bond center~\cite{BJ_Kim_PRL_2008,Kee_Annurev_2016},
and  yields  the hopping to become spin-dependent. In this scenario, the nearest  neighbors   spin-dependent hopping does not violate parity, therefore the even-odd parity mixing of the superconducting gap function for the inter-sublattice Cooper pairings is not expected~\cite{Fischer_PRB_2011,Yanase_PRB_2018}. 
However,  spin-dependent hopping between the next nearest neighbors  (same sublattices) violates parity, and   the staggered spin-orbit interaction has a profound effect on the texture of superconducting triplet $\bd$-vector~\cite{Yip_Annurev_2014}.
To such a degree, still one of the main question 
is  the nodal structure of the order parameter and the precise determination of the momentum dependence of superconducting gap function. This can be investigated by
various experimental methods 
 such as specific heat and thermal transport measurements~\cite{Matsuda_NJP_2006}, angle-resolved photoemission spectroscopy (ARPES)~\cite{Okazaki_Science_2012}, and spectroscopic imaging STM-based quasiparticle interference  (QPI) technique~\cite{Byers_PRL_1993,Wang_PRB_2003,Balatsky:2006aa,Maltseva_PRB_2009,Chuang_Science_2010}. 
The latter is based on the survey of the modulations in electronic  local density of states (LDOS) due to the random  impurities, which
%
in the nodal superconductors is carrying fruitful information about the symmetry of gap function~\cite{JennyHoffman_Science_2002,Hanaguri_Science_2009,Hanaguri_Science_2010,Knolle_PRL_2010,Akbari_PRB_2010,Akbari_PRB_2011,Huang_PRB_2011,Allan_Science_2012,Hoffmann_PRB_2013,Zhang_PRB_2016,Lambert_PRB_2017}. 
Particularly, in the NCS superconductors, the position of nodes are accidental and can not be determined by symmetrical considerations~\cite{Sigrist_AIP_2009,Akbari_EPL_2013,Akbari_EPJ_2013}. 
\\

In this paper we aim to point out that a  similar statement on the nodal structure can be hold for the case of superconducting materials with locally broken inversion symmetry, by considering a  tetragonal crystal  with a non-collinear antiferromagnetic order.
We employ a minimal tight-binding model relevant for  a two-dimensional  NCS structure, which is applicable into a wide variety of materials, such as strongly correlated TMOs and odd-parity magnetic multipole materials.
We study the effect of the first and second neighbor spin-dependent hopping on the electronic band structure of a  locally NCS system.
Moreover, we take into account both inter and intra-sublattice  pairings and try to catch their individual and combined signatures in the electron conductance modulations due to magnetic or nonmagnetic impurities, probed by STM. We conclude that applying  the QPI approach  to the system with locally broken inversion symmetry would be  very informative.

\begin{figure}[t]
	\begin{center}
		\hspace{-0.1cm}
		\includegraphics[width=0.85 \linewidth]{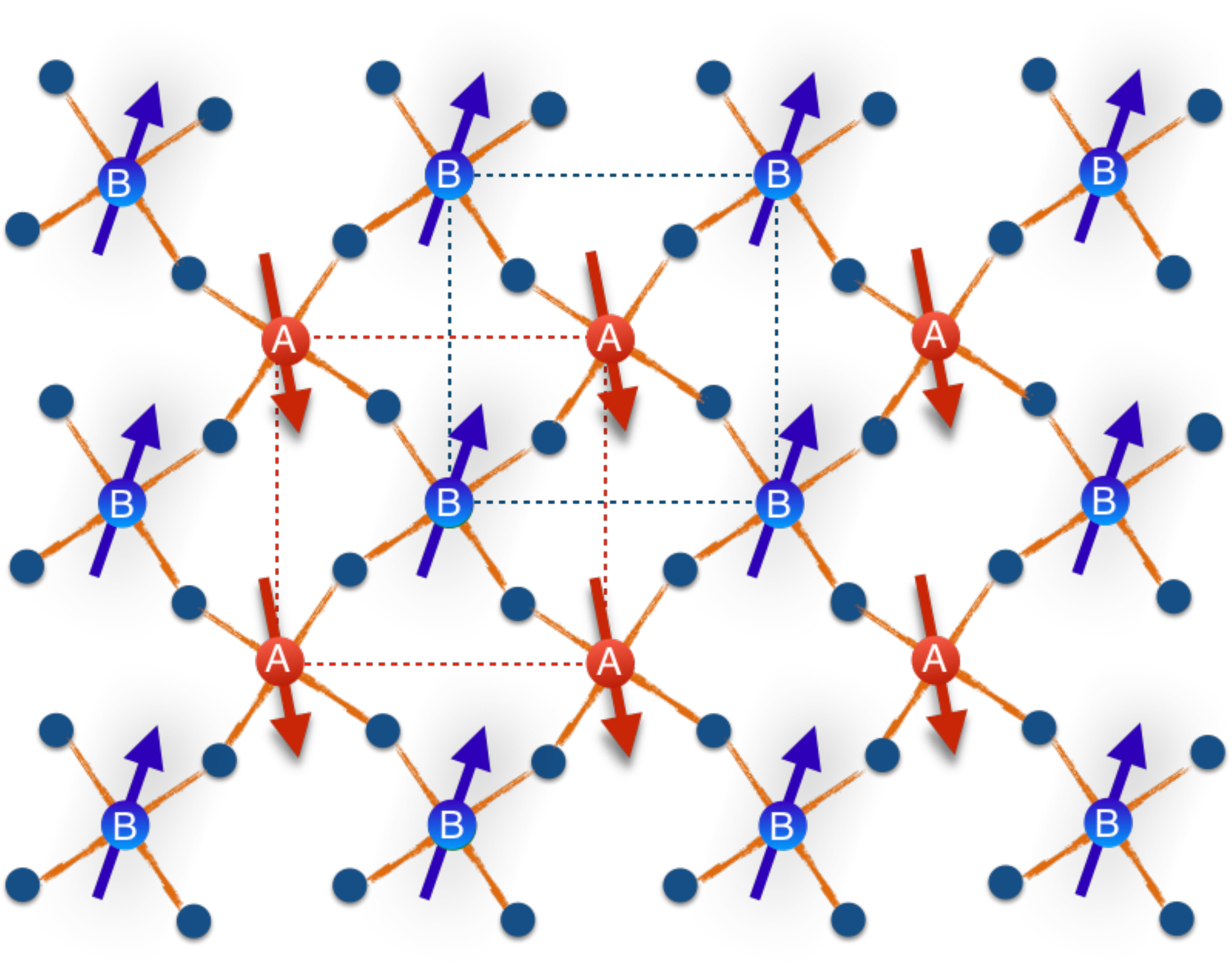} 
	\end{center}
	\vspace{-0.3cm}
	\caption{
		The unit cell of the layered TMO provskite (dashed lines). 
		  A (red) and B (blue) circles denote TM atoms with different magnetic moments and non-collinear antiferromagnetic order along with a partial ferromagnetic moment.
		The dark-blue circles are oxygen atoms.
		The deviation of TM-O-TM bonds from $180^{\circ}$ breaks the local inversion symmetry. 
			}
	\label{Fig:Unit_Cell}
\end{figure}
%

\section{Description of the Model} 
We consider a nonsymmorphic locally NCS system with sublattice degree of freedom.
The lattice consists of a two-dimensional XO$_2$ (X=TM) layer, with the crystal structure illustrated schematically  in Fig.~\ref{Fig:Unit_Cell}.
This figure depicts a noncollinear antiferromagnetic (AFM) structure, which obviously breaks time-reversal symmetry (TRS).
Here, we assume that TRS remains untouched in the system.
This structure belongs to the globally centrosymmetric and nonsymmorphic I$4^{}_{1}/acd$ space group~\cite{Hsieh_PRL_2014}, which is different with I$4/mmm$ fully symmetric space group, because of in-plane rotation of octahedra~\cite{Crawford_PRB_1994}.
The recent studies by single-crystal
neutron diffraction revealed the reduction of space group into I$4^{}_{1}/a$~\cite{Hsieh_PRL_2014,Cao_PRB_2015}. 
Thus, the local site symmetry reduces to D$^{}_{2d}$ due to the lack of local inversion symmetry~\cite{Yanase_PRB_2018}.
We restrict ourselves to the single band Hubbard model, applicable to the canted Mott/spin-orbit-Mott insulating systems~\cite{BJ_Kim_PRL_2008,Moon:2008aa,Yu_PRB_2009,Senthil_PRL_2011,Witczak-Krempa_Annurev_2014,Kee_Annurev_2016,BJ_Kim_Annurev_2019}.
In this picture, the non-interacting Hamiltonian of the model in real space is obtained by
%
\bea
\begin{aligned}
	{\cal H}^{}_{0}
	=&
	-
	\mu
	-
	\!\!\!
	\sum_{ \bra ij\ket_{_1}, \sigma}
	\Big[
	t^{}_1
	[
	a^{\dagger}_{i\sigma} b^{}_{j\sigma}
	+
	b^{\dagger}_{i\sigma} a^{}_{j\sigma}
	]
	\\
	&
	\hspace{1.75cm}
	+
	\sum_{\sigma'}
	{\mi}
	t'^{}_1
	{\sigma}^z_{\sigma\sigma'}
	[
	a^{\dagger}_{i\sigma}
	b^{}_{j\sigma'}
	-
	b^{\dagger}_{i\sigma}
	a^{}_{j\sigma'}
	]
	\Big]
	\\
	&
	-
	\sum_{ \bra ij\ket_{_2}, \sigma}
	\Big[
	t^{}_2
	[
	a^{\dagger}_{i\sigma}
	a^{}_{j\sigma}
	+
	b^{\dagger}_{i\sigma}
	b^{}_{j\sigma}
	]
	\\
	&
	\hspace{1.15cm}
	+
	\sum_{\sigma'}
	{\mi}
	t'^{}_2
	({\boldsymbol{\sigma}} \!
	\times \hat{\br}^{}_{ij})_{\sigma\sigma'}^z
	[
	a^{\dagger}_{i\sigma}
	a^{}_{j\sigma'}
	-
	b^{\dagger}_{i\sigma}
	b^{}_{j\sigma'}
	]
	\Big].
	\hspace{0.0cm}
\end{aligned}
\label{Eq:real_Ham}
\eea
%
Here the antiferromagnet picture is  characterised by operators 
 $a^{\dagger}_{i\sigma}$ and $b^{\dagger}_{i\sigma}$ that create an electron  with spin $\sigma$ at $i$th site of the magnetic sublattices A and B, respectively. 
Also, $t^{}_{1}$ ($t'^{}_{1}$) and $t^{}_{2}$ ($t'^{}_{2}$) are representatives of the spin-independent (dependent) inter and intra-sublattice hopping integrals, respectively.
The spin-dependent terms are arising from locally broken of inversion symmetry~\cite{Kee_Annurev_2016}, and their magnitudes  are controlled  by the  angle of X-O-X bonds~\cite{Yu_PRB_2009}. 
\\

In this picture, the total Hamiltonian of the model is represented by
%
\begin{align}
{\cal H}=
{\cal H}^{}_{0}
+
{\cal H}^{}_{\rm int},
\end{align}
%
where ${\cal H}^{}_{0}$ shows the non-interacting part of Hamiltonian in momentum space.
It is described by creation field operator
$\Phi^{\dagger}_{\bk}=(a^{\dagger}_{\bk\uparrow},a^{\dagger}_{\bk\downarrow},b^{\dagger}_{\bk\uparrow},b^{\dagger}_{\bk\downarrow})$,
defined  by
%
\begin{align}
{\cal H}^{}_{0}=
\sum_{\bk}
\Phi^{\dagger}_{\bk}
\hat{h}^{}_{\bk}
\Phi^{}_{\bk}.
\end{align}
%
Here $\hat{h}^{}_{\bk}$, the corresponding representation matrix,  is expressed  by
\bea
\bl
\hat{h}^{}_{\bk}
\!\!=\!\!
\begin{bmatrix}
\epsilon^{}_{2\bk}-\mu
&
{\rm g}^{x}_{2\bk}
\!-\!{\mi} {\rm g}^{y}_{2\bk}
&
\epsilon^{}_{1\bk}-{\mi}{\rm g}^{}_{1\bk}
&
0
\\
{\rm g}^{x}_{2\bk}+{\mi} {\rm g}^{y}_{2\bk}
&
\epsilon^{}_{2\bk}-\mu
&
0
&
\epsilon^{}_{1\bk}+{\mi}{\rm g}^{}_{1\bk}
\\
\epsilon^{}_{1\bk}+{\mi}{\rm g}^{}_{1\bk}
&
0
&
\epsilon^{}_{2\bk}
\mu
&
-
{\rm g}^{x}_{2\bk}+{\mi} {\rm g}^{y}_{2\bk}
\\
0
&
\epsilon^{}_{1\bk}
\!-\!
{\mi}{\rm g}^{}_{1\bk}
&
-
{\rm g}^{x}_{2\bk}
\!-\!
{\mi} {\rm g}^{y}_{2\bk}
&
\epsilon^{}_{2\bk}-\mu
\end{bmatrix},
\no
\el
\eea
%
where 
%
\begin{align}
\begin{aligned}
\epsilon^{}_{1\bk}
&=
-4t^{}_1 \cos(k_x/\sqrt{2})\cos(k_y/\sqrt{2}),
\\
\epsilon^{}_{2\bk}
&=
-4t^{}_2[\cos(k_x\sqrt{2})+\cos(k_y\sqrt{2})],
\end{aligned}
\label{Eq:1st_2nd_independent}
\end{align}
%
are the energy dispersions originated from the first-neighbor (inter-sublattice) and second-neighbor (intra-sublattice) spin-independent hopping, respectively.
The mathematical difference between the energy dispersion in Eq.~(\ref{Eq:1st_2nd_independent}) is resulted from the $45^{\circ}$ rotation of inter-sublattice hopping vectors with respect to the intra-sublattice ones.
%
\begin{figure}[t]
	\begin{center}
		\hspace{-0.15cm}
		\includegraphics[width=0.99 \linewidth]{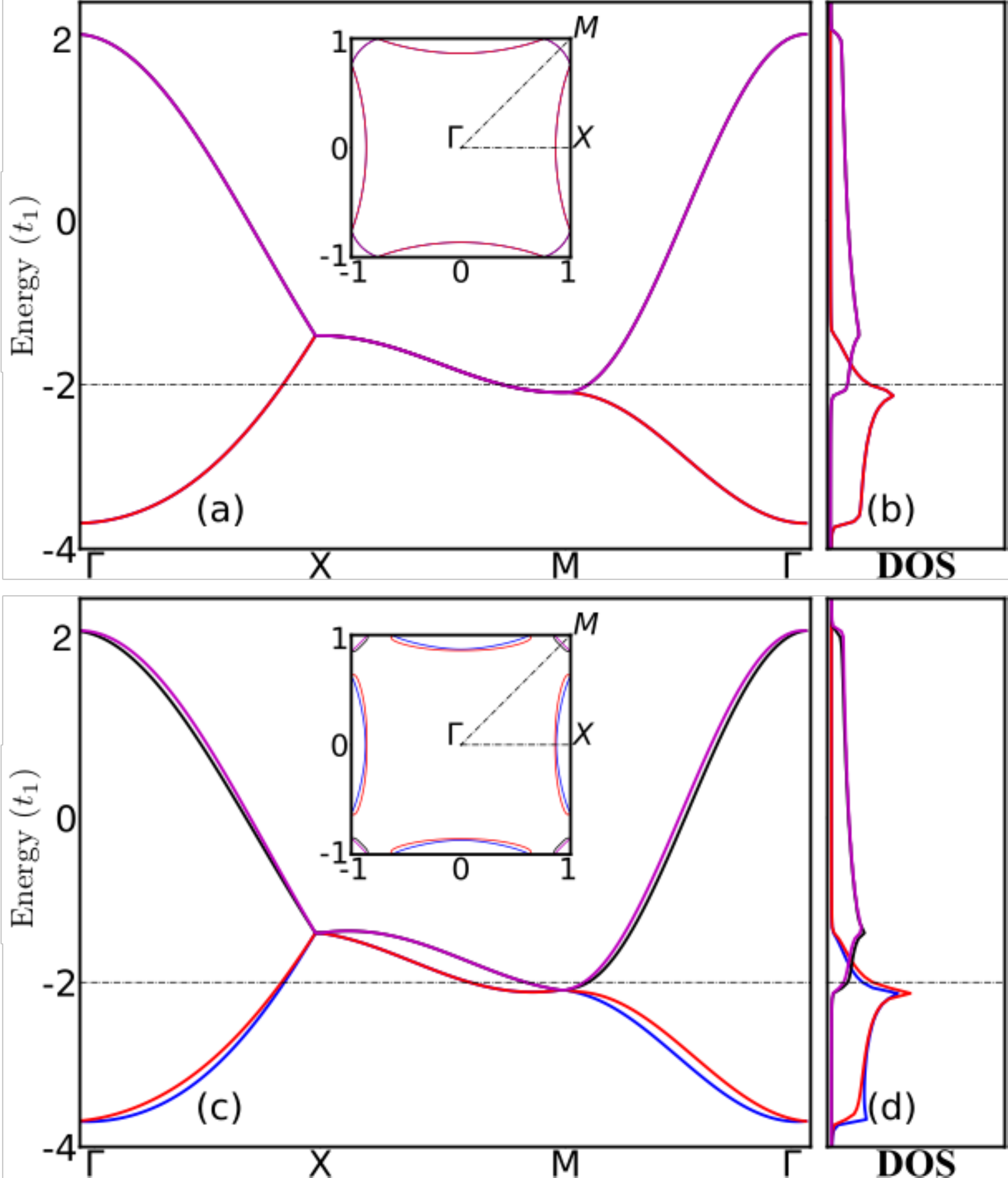} 
	\end{center}
			\vspace{-0.2cm}
	\caption{
		%
		The band dispersion along $\Gamma$XM$\Gamma$ path  
		for  $t'^{}_1=0.5t^{}_1$, $t^{}_2=0.26t^{}_1$, and for two different cases of SQ-SOC with $t'^{}_2=0$  (upper panel), and mixed SQ-SOC and AQ-SOC with $t'^{}_2=0.2t^{}_1$ (lower panel).
		Inset of (a) shows the Fermi surface with a two-fold spin degeneracy, and the inset of  (c) shows that how the Fermi surface has a tiny spin-orbit splitting as a result of mixing the SQ-SOC and AQ-SOC. 
		The band resolved density of states are shown in (b and d), where  (b)/(c)  illustrates the spin degeneracy/the band splitting.
	}
	\label{Fig:FS_DOS}
\end{figure}
%
%
Furthermore,
%
\begin{equation}
\bg^{}_{1\bk}
={\rm g}^{}_{1\bk}\; \hat{z}
=4t'^{}_1\cos(k_x/\sqrt{2})\cos(k_y/\sqrt{2})
\;
\hat{z}
\label{Eq:1st_dependent}
\end{equation}
%
corresponds to the first-neighbor (inter-sublattice) spin-dependent hopping,
and generates a symmetric quasi-SOC (SQ-SOC). 
One should notice that $\bg^{}_{1\bk}$  is an even vector under parity, and therefore 
 never results in a  mixed even-odd parity in inter-sublattice Cooper pairing. 
On top of that, taking into account the spin-dependent intra-sublattice hopping leads to a Rashba-type antisymmetric quasi-SOC (AQ-SOC),
which can be described by an odd $\bg$-vector, such as
%
\begin{equation}
\bg^{}_{2\bk}
\!
=
2t'^{}_2
\Big[
\sin(k_y\sqrt{2})\hat{x}-\sin(k_x\sqrt{2})\hat{y}
\Big].
\label{Eq:AQ-SOC_g_vector}
\end{equation}
%
This term is originated from the combination of the IrO$_6$ octahedra rotation and stacking structure along the c-axis, which breaks the mirror symmetry concerning the ab-plane.
The combination of the IrO$^{}_6$ rotation with the stacking structure of the
2D layers along the c-axis breaks the mirror symmetry with regard to the ab-plane and results in the spin-dependent intra-sublattice term~\cite{Yanase_PRL_2017}. 
Moreover, the crinkling of the lattice by displacing the A (B) sublattice in the $\hat{z}$ ($-\hat{z}$) direction allows for a second-neighbor (intra-sublattice) spin-dependent
hopping~\cite{Kane_PRL_2015}.
In contrast with SQ-SOC, the AQ-SOC violates parity on each sublattice, hence a parity mixing takes place for intra-sublattice Cooper pairs~\cite{Yanase_PRL_2017}.
\\

Diagonalizing the Hamiltonian gives us four effective bands, whose energy dispersion is given by
%
\begin{align}
\varepsilon^{}_{\bk,\pm\pm}=
-\mu+\epsilon^{}_{2\bk}
\pm
\sqrt{
	\epsilon^{2}_{1\bk}
	+
	\Big(
	|\bg^{}_{1\bk}|
	\pm
	\:
	|\bg^{}_{2\bk}|
	\Big)^2_{}
}.
\label{Eq:Normal_Energy}
\end{align}
%
%
In this configuration, the band filling is defined as the number of electrons per unit cell and expressed by $\langle n \rangle=\langle n_0\rangle+2\delta$, where $\delta$ corresponds  the level of doping  and   $\langle n_0 \rangle=2$ represents a half-filling.
The band structure 
 for the electron-doped case with $\delta=0.2$, is depicted in Fig.~\ref{Fig:FS_DOS} along the  $\Gamma$XM$\Gamma$ path. 
Excluding the intra-sublattice hopping results in two two-fold degenerate bands without spin splitting, as it is shown in Fig.~\ref{Fig:FS_DOS}(a).
The inset represents the structure of Fermi surface.
Fig.~\ref{Fig:FS_DOS}(b) illustrates density of states (DOS) in the absence of intra-sublattice spin-dependent hopping.
From Eq.~(\ref{Eq:Normal_Energy}), it is   obviously seen that the splitting occurs when  both SQ-SOC and AQ-SOC are present simultaneously.
Including both quasi-SOCs, Fig.~\ref{Fig:FS_DOS}(c) represents the band structure and Fermi surface (inset)  for high symmetric path with generating a tiny spin splitting.
Fig.~\ref{Fig:FS_DOS}(d) depicts the band resolved electronic DOS, which obviously reveals this spin splitting.
In addition, the interacting part of the non-superconducting Hamiltonian in momentum space is obtained by the on-site Hubbard term as
%
\begin{align}
{\cal H}^{}_{\rm int}
\!
=
\!
\frac{U}{2}
\!\!
\sum_{\bq\bk\bk'\sigma}
\!\!\!
\Big[
a^{\dagger}_{\bk+\bq\sigma}
a^{\dagger}_{\bk'-\bq\bar{\sigma}}
a^{}_{\bk'\bar{\sigma}}
a^{}_{\bk\sigma}
\!
+
b^{\dagger}_{\bk+\bq\sigma}
b^{\dagger}_{\bk'-\bq\bar{\sigma}}
b^{}_{\bk'\bar{\sigma}}
b^{}_{\bk\sigma}
\!
\Big].
\label{Eq:Hubbard}
\end{align}
%
%
\begin{figure}[t]
	\begin{center}
		\hspace{-0.1cm}
		\includegraphics[width=0.85 \linewidth]{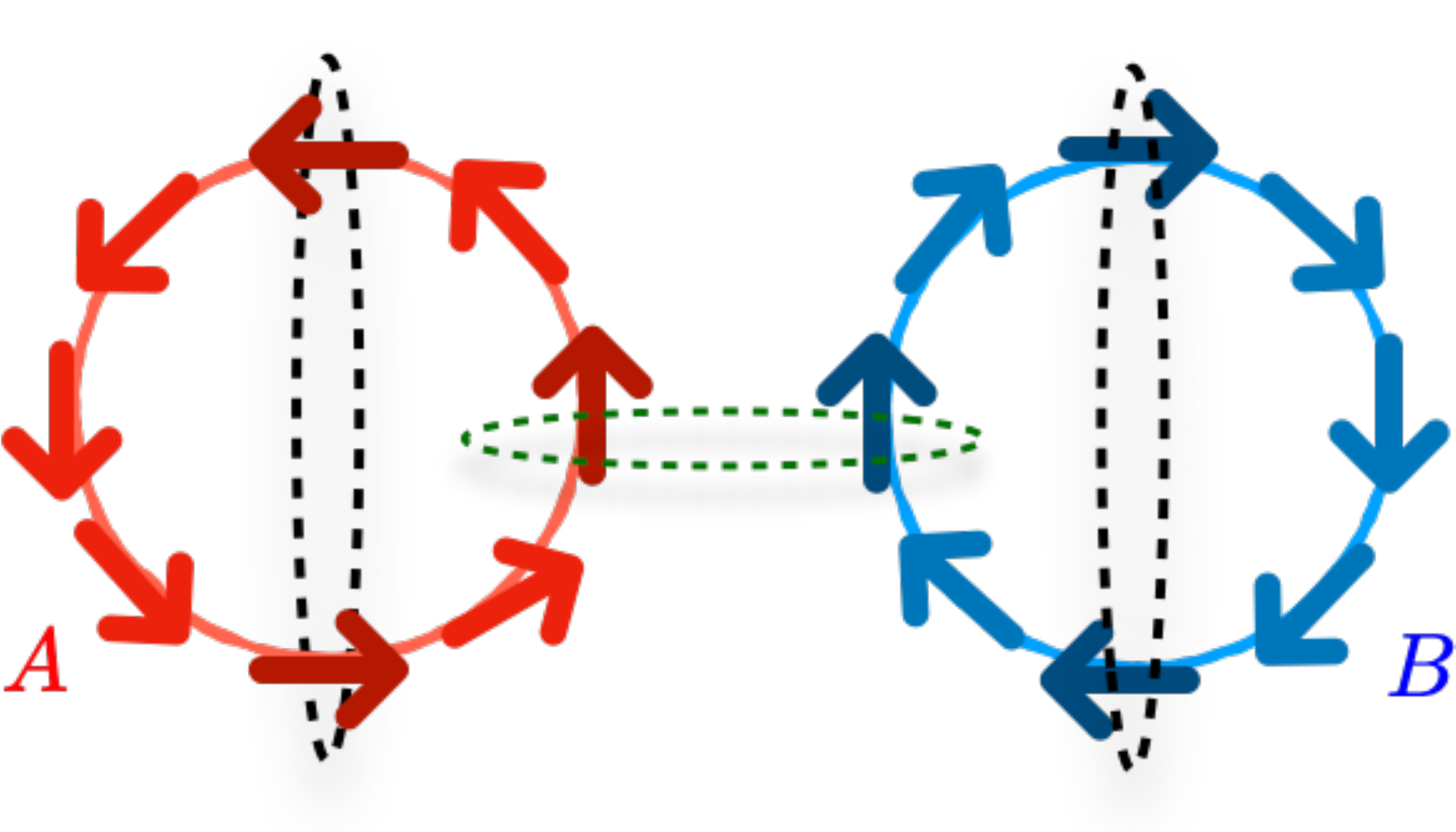} 
	\end{center}
				\vspace{-0.4cm}
	\caption{
		Schematic  representation of Cooper pairing for inter (green dashed line) and intra-sublattice (black dashed line) pairings 
		between two fermions with opposite sign  momenta ($\bk$ and $-\bk$)~\cite{Yanase_PRB_2018}.
		Different magnetization on sublattices makes opposite spin textures on Fermi surfaces.
		Note: the center of  Fermi surface of B-sublattice is shifted for better representation.
	}
	\vspace{-0cm}
	\label{Fig:Pairing}
\end{figure}
%

To discuss the superconducting state, we employ  Hamiltonian within the Nambu basis $\Psi^{\dagger}_{\bk}=(\Phi^{\dagger}_{\bk},\Phi^{}_{-\bk})$, expressed as
%
\begin{equation}
{\cal H}^{}_{\rm BdG}=
\sum_{\bk}
\Psi^{\dagger}_{\bk}
\check{\cal H}^{}_{\rm SC}(\bk)
\Psi^{}_{\bk},
\end{equation}
%
with
%
\begin{equation}
\check{\cal H}^{}_{\rm SC}(\bk)=
\begin{bmatrix}
\hat{h}^{}_{\bk}
&
\hat{\Delta}^{}_{\bk}
\\
\hat{\Delta}^{\dagger}_{\bk}
&
-\hat{h}^{*}_{-\bk}
\end{bmatrix},
\end{equation}
%
where, the  superconducting gap matrix in sublattice-spin basis is given by
\begin{align}
\begin{aligned}
\hat{\Delta}^{}_{\bk}=
\tilde{\Delta}^{\rm intra}_{\bk}
\;
\zeta^{}_{0}
+
\tilde{\Delta}^{\rm inter}_{\bk}
\;
\zeta^{}_{x},
\end{aligned}
\end{align}
%
and $\zeta^{}_{i}(i=0,x)$ denote the Pauli matrices in the sublattice space.
In addition, $\tilde{\Delta}^{\rm intra}_{\bk}$ and $\tilde{\Delta}^{\rm inter}_{\bk}$ are defining the intra and inter-sublattice superconducting gap functions, respectively.
It is worthwhile to mention that both gap functions have the well-known form of superconducting order parameter in spin basis as following
\begin{align}
\begin{aligned}
\tilde{\Delta}^{\rm inter}_{\bk}
&=
{\mi}[
\bd^{\rm inter}_{\bk} \cdot \boldsymbol{\sigma}
]\sigma^{}_y,
\\
\tilde{\Delta}^{\rm intra}_{\bk}
&=
{\mi}[
\psi^{\rm intra}_{\bk}\sigma^{}_0
+
\bd^{\rm intra}_{\bk} \cdot \boldsymbol{\sigma}
]\sigma^{}_y.
\end{aligned}
\end{align}
%
In the above equations, $\psi^{}_{\bk}$ and $\bd^{}_{\bk}$ are the singlet and triplet pairing functions, respectively.
They correspond  to $B^{}_{1g}$ and $B^{}_{2g}$ irreducible representations of $D^{}_{4h}$, considering the strength of electron correlation and quasi-SOCs.
The inter- and intra-sublattice pairings for $B^{}_{1g}$ irreducible representation are given by
%
\begin{align}
\begin{aligned}
{\bd}^{\rm inter}_{\bk}
&=
\Delta^{\rm inter}_{\rm t}
\Big[
\sin(\frac{k_x}{\sqrt{2}})
\cos(\frac{k_y}{\sqrt{2}})
\hat{x}
+
\cos(\frac{k_x}{\sqrt{2}})
\sin(\frac{k_y}{\sqrt{2}})
\hat{y}
\Big],
\\
\psi^{\rm intra}_{\bk}
&=
\Delta^{\rm intra}_{\rm s}
\Big[
\sin(k_x\sqrt{2}) \sin(k_y\sqrt{2})
\Big],
\\
{\bd}^{\rm intra}_{\bk}
&=
\Delta^{\rm intra}_{\rm t}
\Big[
\sin(k_x\sqrt{2})\hat{x}
+
\sin(k_y\sqrt{2})\hat{y}
\Big].
\end{aligned}
\end{align}
%
For B$^{}_{2g}$ representation, these components  are represented by
%
\begin{align}
\begin{aligned}
{\bd}^{\rm inter}_{\bk}
&=
\Delta^{\rm inter}_{\rm t}
\Big[
\cos(\frac{k_x}{\sqrt{2}})
\sin(\frac{k_y}{\sqrt{2}})
\hat{x}
-
\sin(\frac{k_x}{\sqrt{2}})
\cos(\frac{k_y}{\sqrt{2}})
\hat{y}
\Big],
\\
\psi^{\rm intra}_{\bk}
&=
\Delta^{\rm intra}_{\rm s}
\Big[
\cos(k_x\sqrt{2}) - \cos(k_y\sqrt{2})
\Big],
\\
{\bd}^{\rm intra}_{\bk}
&=
\Delta^{\rm intra}_{\rm t}
\Big[
\sin(k_y\sqrt{2})\hat{x}
-
\sin(k_x\sqrt{2})\hat{y}
\Big].
\end{aligned}
\end{align}
%

%
The singlet component of $B^{}_{2g}$ state is reminiscent of the famous $d^{}_{x^2-y2}$-wave spin-singlet superconducting state that is well-studied in the single-sublattice Hubbard model.
However, its counterpart in $B^{}_{1g}$ irreducible representation corresponds to the spin-singlet $d^{}_{xy}$-wave state\cite{Sigrist_AIP_2009,Bauer_Sigrist_NCS_Book_2012}.
Fig.~\ref{Fig:Pairing} schematically shows that the superconducting order parameter is strongly sublattice dependent.
For intra-sublattice pairing, since the spin texture of both sites $i$ and $j$ are the same, then spin orientations of two electrons with momenta $\bk$ and $-\bk$ are opposite.
Therefore, both spin singlet and triplet Cooper pairs with $S^{}_z=0$ can be realized, and because of violation of parity for the same sublattices, the superconducting gap have mixed even-odd parity.
On the other hand, in the case of inter-sublattice pairing, the spins of electrons on sites $i$ and $j$ with opposite momenta align in the same direction.
Thus, only spin-triplet odd-parity Cooper pairs with $S^{}_z=1$ are allowed to emerge.
It is  shown that inter-sublattice pairing corresponds to $l=1$ with pure triplet superconductivity.
However, for intra-sublattice pairing the relevant form of gap function can be either 
$d^{}_{xy}+p$ or $d^{}_{x^2-y^2}+p$, corresponding to $B^{}_{1g}$ and $B^{}_{2g}$ irreducible representations, respectively~\cite{Yanase_PRB_2018}.
\\

\begin{figure}[t]
	\begin{center}
		\includegraphics[width=1.0 \linewidth]{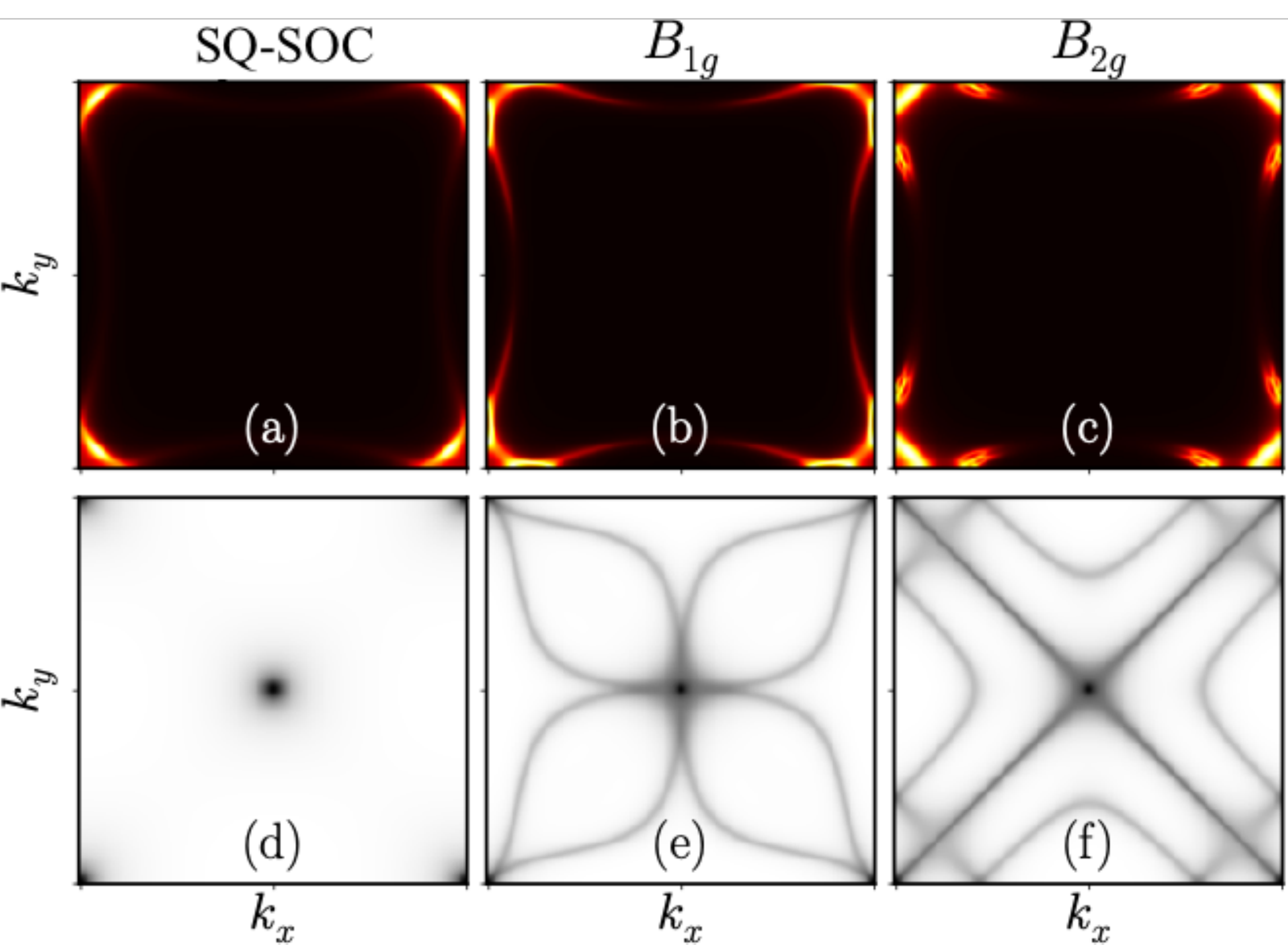} 
	\end{center}
				\vspace{-0.2cm}
	\caption{
		The spectral function of the superconducting state at $\omega=0.05 t^{}_1$, for a locally noncentrosymmetric system 
		corresponding to the tight binding parameters  of the Fig.~\ref{Fig:FS_DOS}.
		(a)  for pure triplet pairing excluding intra-sublattice hopping,  and
		(b  and c) in the presence of both inter- and intra-sublattice hopping in $B^{}_{1g}$ and $B^{}_{1g}$ irreducible representations, respectivlely.
		(d-f) show the nodal structure of the superconducting gap.
			The amplitudes of superconducting gap components are set as 
			$\Delta^{\rm intra}_{\rm s}=\Delta^{\rm inter}_{\rm t}=2\Delta^{\rm intra}_{\rm t}=0.3t_{1}$.
		The plots ranges are   ($-\pi/\sqrt{2},\pi /\sqrt{2}$). 
	}
	\label{Fig:Spec}
\end{figure}

In Fig.~\ref{Fig:Spec}(a), we represent the spectral functions of the superconducting state for
the $B^{}_{2g}$ irreducible representation with inter-sublattice hopping ($t^{}_2=t'^{}_2=0$) and
 pure triplet pairing, at 
energy $\omega=0.05 t^{}_1$.
Fig.~\ref{Fig:Spec}(b and c) portray the superconducting spectral function 
considering the contributions of both inter- and intra-sublattice hopping  (SQ-SOC and AQ-SOC) within $B^{}_{1g}$ and $B^{}_{2g}$ irreducible representations, respectively.
These results are consistent with the formation of Fermi arcs at low temperature and small amount of electron doping reported for Sr$_2$IrO$_4$~\cite{BJ_Kim_Science_2014}, and  describe a nodal structure in the spectral function.
 The nodal structure of corresponding superconducting gap  are shown in Figs.~\ref{Fig:Spec}(d-f), which are describing the zeros of order parameters. 
 %
%
%
The spectral functions within $B^{}_{1g}$ and $B^{}_{2g}$ irreducible representations of superconducting gap are completely dominated by the structure 
of intra-sublattice pairing, which is expected to be reflected in QPI patterns.
 \\
 
 %
%
%
 
 %
\begin{figure}[t]
	\begin{center}
		\hspace{-0.1cm}
		\includegraphics[width=0.98 \linewidth]{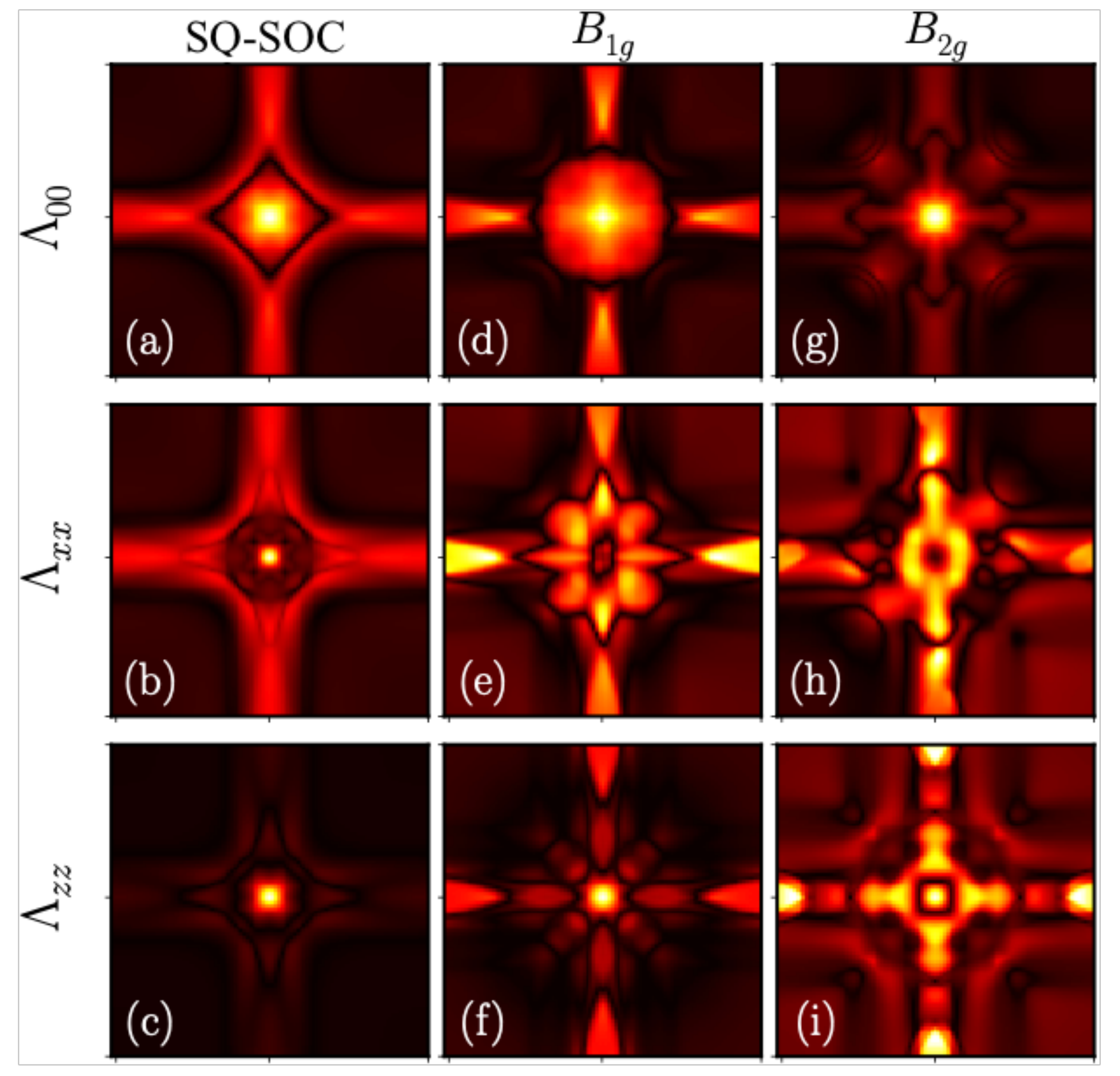} 
		\\
		\hspace{0.4cm}
		\includegraphics[width=0.4 \linewidth]{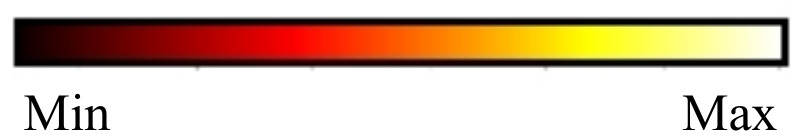}
	\end{center}
			\vspace{-0.2cm}
	\caption{
		The QPI pattern of a locally noncentrosymmetric system in charge ($\Lambda^{}_{00}$) and spin ($\Lambda^{}_{ii}$) channels:  (a-c)  for pure inter-sublattice triplet pairing, (d-f)  including inter- and intra-sublattice pairings within $B^{}_{1g}$ irreducible representation, and (g-i) superposition of both inter and intra-sublattice pairings in $B^{}_{2g}$ irreducible representation.
		The energy is set at $\omega=0.05t_1$, and plot ranges are     ($-\pi/\sqrt{2},\pi /\sqrt{2}$) in $({{q}}_{x},{{q}}_{y})$-plane. 
		Note: The component $\Lambda^{}_{yy}$ can be obtained by performing a $90^{\circ}_{}$ rotation on $\Lambda^{}_{yy}$, 
		as a result of broken spin rotational symmetry in $xy$-plane.
	}
	\label{Fig:QPI_patterns}
\end{figure}
%
%

%
To specifically study the QPI image,
 we consider the  total scattering  Hamiltonian  from a impurity  as
%
\bea
{\cal H}^{}_{\rm imp}=
\sum_{\bk\bq\alpha}
V^{}_{\alpha}(\bq)
S^{}_{\alpha}
\Psi^{\dagger}_{\bk}
\check{\varrho}^{}_{\alpha}
\Psi^{}_{\bk},
\eea
%
where $\lbrace \check{\varrho}^{}_{\alpha}\rbrace=(\check{\varrho}^{}_0,\check{\boldsymbol\varrho})=(\tau^{}_{z}\hat{\rho}^{}_{0},\tau^{}_{0}\hat{\rho}^{}_{x},\tau^{}_{z}\hat{\rho}^{}_{y},\tau^{}_{0}\hat{\rho}^{}_{z})$ are $8\times 8$ matrices, and $\hat{\rho}^{}_{\alpha}=\zeta^{}_{0}\sigma^{}_{\alpha}$.
Besides, the $2\times 2$ matrices $\tau^{}_{i}(i=0,x,y,z)$ are indicating the Pauli matrices in the particle-hole spaces.
Moreover, the parameter $\lbrace S^{}_{\alpha} \rbrace=(1,\bS)$ stands for nonmagnetic $(\alpha=0)$ impurity, and magnetic $(\alpha=x,y,z)$ impurity of spin $\bS$.
The scattering centers lead to the modulation of LDOS in both charge and spin channels.
The elements of QPI intensity are given by~\cite{Akbari_EPL_2013,Akbari_EPJ_2013}
\be
\label{QPI}
\Lambda^{}_{\alpha\beta}
(\bq,\omega)=-\frac{1}{\pi}
{\rm Im}
\Big[
\Upsilon^{}_{\alpha\beta}
(\bq,{\mi}\omega^{}_n)
\Big]_{{\mi}\omega^{}_n\rightarrow \omega+{\mi} 0^+}
,
\ee
%
in which ${\mi}\omega^{}_n=2{\mi}n\pi T$ is bosonic Matsubara frequency at temperature $T$, and
\begin{align}
\Upsilon^{}_{\alpha\beta}
(\bq,{\mi}\omega^{}_n)
=
\frac{1}{N}
\sum_{\bk}
{\rm Tr}
\Big[
\check{\varrho}^{}_{\alpha}
\hat{ G}(\bk,{\mi}\omega)
\check{\varrho}^{}_{\beta}
\hat{ G}(\bq',{\mi}\omega^{}_n)
\Big],
\no
\end{align}
%
where we set $\bq'=\bk+\bq$, and $N$ is the number of grid points.
The  Green's function $\hat{ G}(\bq,{\mi}\omega)$ is obtained by
%
\be
\bl
\hat{ G}(\bk,{\mi}\omega^{}_n)=
\Big[
	{\mi}\omega^{}_n \check{\mathbb{I}}
-
\check{\cal H}^{}_{\rm SC}(\bk)
\Big]^{-1}
,
\el
\ee
%
where  $\check{\mathbb{I}}$ denotes an $8\times 8$ identity matrix.
The Green's function in  Nambu space can be defined by
%
\begin{equation}
\hat{ G}(\bk,{\mi}\omega^{}_n)=
\begin{bmatrix}
\hat{\cal G}(\bk,{\mi}\omega^{}_n)
&
\hat{\cal F}(\bk,{\mi}\omega^{}_n)
\\
\hat{\cal F}^{\dagger}_{}(\bk,{\mi}\omega^{}_n)
&
\hat{\cal G}^{\intercal}_{}(-\bk,-{\mi}\omega^{}_n)
\end{bmatrix},
\no
\end{equation}
%
that $\hat{\cal G}(\bk,{\mi}\omega^{}_n)$ and $\hat{\cal F}(\bk,{\mi}\omega^{}_n)$
are the normal and anomalous Matsubara Green's functions in sublattice-spin basis, respectively.
Consequently, using Eq.~(\ref{QPI}),  the final form of the  QPI elements  in both charge ($\Lambda^{}_{00}$) and spin ($\Lambda^{}_{ii}$) channels are obtained from 
\begin{align}
\begin{aligned}
&
\Upsilon^{}_{\alpha\beta}
(\bq,{\mi}\omega^{}_n)
=
\frac{1}{N}
\sum_{\bk,uv}
\sum_{\lbrace s^{}_i \rbrace}
\;
\hat{\rho}^{s^{}_1\!s^{}_3}_{\alpha}
\;
\hat{\rho}^{s^{}_2\!s^{}_4}_{\beta}
\;
\times
\\
&
\Big[
\hat{\cal G}^{uv}_{s^{}_3\!s^{}_2}(\bk,\!{\mi}\omega_n\!)
\hat{\cal G}^{vu}_{s^{}_4\!s^{}_1}(\bq'\!,{\mi}\omega_n\!)
\!
+
\!
\eta
\hat{\cal F}^{uv}_{s^{}_3\!s^{}_2}(\bk,{\mi}\omega_n\!)
\hat{\cal F}^{ vu\dagger}_{s^{}_4\!s^{}_1}(\bq' \!,{\mi}\omega_n\!)
\Big],
\no
\end{aligned}
\label{Eq:QPI_equation}
\end{align}
%
where $u,v$  are running over the sublattice degree of freedom (A and B), and 
$\eta=1$ ($\eta=-1$) corresponds to $\alpha=x,~z$ ($\alpha=0,~y$).
Although the off-diagonal QPI patterns can be produced as a result of AQ-SOC, here, for simplicity, we only investigate the diagonal elements, $\Lambda^{}_{\alpha\alpha}$, and ignore the spin-polarized patterns generated by nonmagnetic impurity.
Fig.~\ref{Fig:QPI_patterns} shows the numerical results of the diagonal elements of QPI matrix for both nonmagnetic and magnetic impurities.
As can be seen in Fig.~\ref{Fig:QPI_patterns}(a-c), in the absence of  AQ-SOC,  a locally noncentrosymmetric system with pure triplet Cooper pairings preserving in-plane spin rotational symmetry.
It is easily seen that the QPI patterns reflect the topology of normal Fermi surface, since the superconducting order parameter is fully gaped.
Moreover, all diagonal elements of QPI patterns maintain the $C^{}_{4v}$ symmetry, because SQ-SOC has only out-of-plane vector.
Figs.~\ref{Fig:QPI_patterns}(d-f) correspond to 
 the admixture of inter- and intra-sublattice superconducting pairs in $B^{}_{1g}$ irreducible representation.
It can be clearly observed that the singlet $d^{}_{xy}$-wave singlet pairing is dominant over the QPI patterns.
Finally, Figs.~\ref{Fig:QPI_patterns}(g-i) denote the QPI patterns for a system considering both inter- and intra-sublattice pairings in $B^{}_{2g}$ irreducible representation, which clearly reflects the nodal structure of gap function, especially singlet $d^{}_{x^2-y^2}$-wave pairing.
It should be noted that the in-plane components of spin-polarized QPI patterns $\Lambda^{}_{xx}$, and $\Lambda^{}_{yy}$ have a $C_{2v}$ symmetry,
 and they are equivalent together under a $90^{\circ}$ rotation around the center of Brillouin zone.
Moreover, the peaks in the QPI patterns are obviously coming from the scattering events between the states with the highest DOS in the spectral function.
\\

\section{Conclusion}
We study the effects of inter and intra-sublattice  (first and second neighbors, made up of both spin-dependent/independent) hopping amplitudes on the electronic structure of a locally noncentrosymmetric system.
We have found that neglecting the effect of intra-sublattice (second neighbor) hopping leads to maintain the two-fold spin degeneracy, which preserves in-plane spin rotational symmetry.
However, including the  intra-sublattice hopping, lifts the spin degeneracy and breaks SU(2) symmetry, similar to the case of a globally NCS. 
Besides, we study  the theory of QPI for a   locally NCS superconductor, in the presence of both magnetic and nonmagnetic  impurities.
%
We show that considering only a inter-sublattice hopping, 
the QPI patterns mainly reflect the structure of normal Fermi surface. 
However, including  the   intra-sublattice hopping, the nodal characteristics of gap function leads to some dominant scattering in the QPI patterns, 
and it causes to breaking of SU(2) symmetry, reducing the symmetry of QPI in spin channels ($\Lambda^{}_{xx}$, and $\Lambda^{}_{yy}$) into $C^{}_{2v}$.
Finally, superposition of both inter and intra-sublattice pairing reveals that the scattering phenomena in the intra-sublattice pairings dominates over the QPI patterns.
\\

%
\section{Acknowledgments}
 We are grateful to G.~Jackeli, and B.J.~Kim for fruitful discussions.
 This work  is supported  through National Research Foundation (NRF) funded by the Ministry of Science of Korea (Grants  No. 2017R1D1A1B03033465 \& No. 2019R1H1A2039733).
 A.A. acknowledges the support of the Max Planck- POSTECH-Hsinchu Center for Complex Phase Materials, and financial support from the National Research Foundation (NRF) funded by the Ministry of Science of Korea (Grant No. 2016K1A4A01922028).
M.\ B. acknowledges the receipt of the grant No. AF-03/18-01 from Abdus Salam International Center for Theoretical Physics, Trieste, Italy.
A.A. acknowledges the National Foundation of Korea  funded by the Ministry of Science, ICT and Future Planning (No. 2016K1A4A4A01922028).
 


\bibliographystyle{epj}
\bibliography{Refs}

\end{document}